\documentclass[useAMS,usenatbib]{mn2e}
\usepackage{mathptmx}
\usepackage{graphicx}
\usepackage{amsmath}

\title[Imprint of Inhomogeneous and Anisotropic Primordial Power Spectrum on CMB Polarization]{Imprint of Inhomogeneous and Anisotropic Primordial Power Spectrum on CMB Polarization}
\author[Rahul Kothari, Shamik Ghosh, Pranati K. Rath, Gopal Kashyap and Pankaj Jain]{Rahul Kothari\thanks{E-mail:
rahulko@iitk.ac.in} 
Shamik Ghosh\thanks{E-mail:
shamik@iitk.ac.in} 
Pranati K. Rath\thanks{E-mail:
pranati@iitk.ac.in} 
Gopal Kashyap\thanks{E-mail:
gopal@iitk.ac.in} 
and Pankaj Jain\thanks{E-mail:
pkjain@iitk.ac.in} \\
Indian Institute of Technology, Kanpur 208016, India\\}
\begin{document}



\maketitle

\label{firstpage}

\begin{abstract}
We consider an inhomogeneous model and independently an anisotropic model of primordial 
power spectrum in order to describe the observed hemispherical anisotropy 
in Cosmic Microwave Background Radiation. This anisotropy can be parametrized in terms
of the dipole modulation model of the temperature field. Both the models lead 
to correlations between spherical harmonic coefficients corresponding to multipoles, $l$ and $l\pm1$. 
We obtain the model parameters by making a fit to TT correlations in CMBR data. Using these parameters we predict the signature of our 
models for correlations among different multipoles for the case of the TE and EE modes.
These predictions can be used to test whether the observed hemispherical anisotropy can be correctly 
described in terms of a primordial power spectrum. Furthermore these may also allow us to distinguish between an inhomogeneous and an anisotropic model.
\end{abstract}


\section{Introduction}

There currently exist several observations which indicate a potential violation
of the Cosmological Principle which states that the Universe is homogeneous and
isotropic on large distance scales. One such observation is the hemispherical 
power  asymmetry. \citep{Eriksen2004, Eriksen2007, Erickcek2008, Hansen2009, 
Hoftuft2009, Paci2013, Planck2013a, Schmidt2013, Akrami2014} in the Cosmic 
Microwave Background Radiation (CMBR) which has 
generated considerable interest in recent years. Parametrization of hemispherical
anisotropy of CMBR temperature field can be obtained by the dipole modulation 
model \citep{Gordon2005, Gordon2007, Prunet2005, Bennett2011} of the following 
form:
\begin{equation}
 \Delta T\left(\vec{n}\right) = f\left(\vec{n}\right)\left(1+A\hat{\lambda}\cdot\vec{n}\right),\label{eq:dipole_mod}
\end{equation}
where $f(\vec{n})$ is an isotropic field, $\hat{\lambda}$ the preferred direction 
and $A$ amplitude of dipole modulation. Both WMAP and PLANCK \citep{Planck2013a}
data confirm this asymmetry signal with high significance. This effect is absent 
at high-$l$ \citep{Donoghue2005,Hanson2009}. The dipole amplitude $A$ and the direction as observed in the 
WMAP five year data are $0.072\pm0.022$ and 
$(l,b)=(224^{\circ},-22^{\circ})\pm24^{\circ}$ respectively \citep{Hoftuft2009}. 
The PLANCK observations are consistent with this result. In order to explain
the observed large scale anisotropy many theoretical models have been proposed
\citep{Berera2004,ACW2007,Boehmer2008,Jaffe2006,Koivisto2006,Land2006,Bridges2007,Campanelli2007,Ghosh2007,Pontezen2007,Koivisto2008,Kahniashvili2008,Carroll2010,Watanabe2010,Chang2013a,Anupam2013a,Anupam2013b,Cai2013,Liu2013,Chang2013b,Mcdonald2014,Ghosh2014}.

In this paper we are interested in the possibility that the dipole modulation might
be related to the primordial power spectrum. A fascinating possibility is that it might 
be related to the early inhomogeneous or anisotropic phase of the expansion of the universe. 
Such a phase is perfectly consistent with the Big Bang paradigm which 
postulates that the Universe acquires isotropy during inflation. This phenomenon has been 
explicitly demonstrated with in the framework of the anisotropic Bianchi Models \citep{Wald1983}. 
It has been found that most of these models become isotropic very quickly during inflation. 
It has been suggested that the anisotropic modes, generated during
this early phase may later re-enter the horizon \citep{Aluri2012,Pranati2013a} and lead to observed signals of
anisotropy. These modes will primarily affect the low $l$ multipoles of the CMBR and hence 
are consistent with the observations.

We consider an inhomogeneous and, independently, an anisotropic model of the primordial
power spectrum. We assume a simple parametrization of the power spectrum in these
models. We expect that such a form may be related to a fundamental model which might
involve an inhomogeneous metric or noncommutative space time or some other model of
physics at very high energy scale. Here we use the fact that the observed anisotropy is a small 
effect and hence we only need to keep its contribution at leading order. We assume a simple functional 
form of the $k$ dependence of this contribution and extract
the parameters of this function by making a fit to the temperature data. These 
parameters are then used to predict the anisotropy in polarization data. In particular
we focus on the  EE  and  TE correlations. The power spectrum is assumed 
to be generated by a scalar inflaton field. Since such a field
does not generate B modes, we don't consider correlations involving these modes.
The signature of hemispherical anisotropy in CMBR polarization power 
spectrum has been considered earlier in \citep{ChangWang2013b,Namjoo2014,Zarei2014}. 
However our emphasis and results are different. In particular we 
relate the observed anisotropy in temperature data to a primordial power
spectrum and use the resulting power in order to predict the
the signature in CMBR polarization. These results are so far not available
in literature.

We next clarify that if we assume homogeneity then it is not possible to
generate a dipolar power spectrum required for an explanation of the 
 hemispherical anisotropy within the  
 standard framework of commutative space-time. However there are indications
that such a power spectrum may be obtained within a non-commutative space-time
\citep{Akofor2008,Akofor2009,Koivisto2011,Groeneboom2011,Pranati2014b,Kothari2015}.   
Here we shall not be concerned with the theoretical subtleties in the 
calculation of this power spectrum and its relationship with the observed CMBR 
correlations. These issues are partially addressed in \citet{Kothari2015}. However considerable more work is 
required before this framework can be properly understood. In any case
the inhomogeneous model provides a theoretically sound description of the observed
hemispherical anisotropy. Hence we expect our results for the
case of an inhomogeneous model to be very reliable.

\section{Theory}
\label{Sec:Theory}
The CMBR temperature fluctuation and polarizations are fields on a sphere, so they
can be decomposed in spherical harmonics
\begin{equation}
 \Delta T(\vec{n})  =\sum_{lm}a_{T,lm}Y_{lm}(\vec{n}),\label{eq:temp_expansion}
\end{equation}

\begin{equation}
 \tilde{E}\left(\vec{n}\right)  =\sqrt{\frac{\left(l+2\right)!}{\left(l-2\right)!}}\sum_{lm}a_{E,lm}Y_{lm}\left(\vec{n}\right)\label{eq:batti_expansion}
\end{equation}
where $a_{X,lm}$ are the spherical harmonic coefficients and $X$ is the index denoting
temperature (T) or E-Mode polarization (E). In terms of primordial density perturbation
$\delta(\vec{k})$, they are defined by the equation:
\begin{equation}
a_{X,lm}=\left(-i\right)^{l}T_{0}4\pi\int d^{3}{k}\delta(\vec{k})\Delta_{X}\left(l,k,\tau_{0}\right)Y_{lm}^{*}\left(\hat{k}\right)\label{eq:alm}
\end{equation}
where $\tau_{0}$ is the present conformal time, $T_{0}$ is the mean CMBR temperature 
and $\Delta_{X}\left(l,k,\tau_{0}\right)$ is the standard transfer function
employed in CAMB. 
Using the standard notation \citep{Zaldarri1997}, it can be written as, 
\begin{equation}
\Delta_{X}\left(l,k,\tau_{0}\right)=\begin{cases}
\int_{0}^{\tau_{0}}d\tau S_{T}^{(S)}\left(k,\tau\right)j_{l}\left(x\right), & X=T\\
\frac{3}{4}\sqrt{\frac{\left(l+2\right)!}{\left(l-2\right)!}}\int_{0}^{\tau_{0}}d\tau g\left(\tau\right)\Pi\left(k,\tau\right)\frac{j_{l}\left(x\right)}{x^{2}}, & X=E.
\end{cases}\label{eq:gen_def_trans_func}
\end{equation}

The two point correlation
function in the presence of dipole modulation term in the temperature field can
be written as:
\begin{equation}
\langle a_{T,lm}a_{T,l'm'}^{*}\rangle=\langle a_{T,lm}a_{T,l'm'}^{*}\rangle_{ iso}+\langle a_{T,lm}a_{T,l'm^{\prime}}^{*}\rangle_{dm},\label{eq:twopointcorr}
\end{equation}
where $\langle a_{lm}a_{l^{\prime}m^{\prime}}^{*}\rangle_{iso}=C_{l}\delta_{ll^{\prime}}\delta_{mm^{\prime}}$.
Here $C_{l}$ is the isotropic angular power spectrum and the second term on right 
is the contribution because of the dipole modulation term in the temperature field. It has been shown \citep{Prunet2005,Pranati2013b} that the dipole modulation leads to a correlation between the $l$ and $l\pm1$ multipoles of the temperature field.

Using Eq. \ref{eq:alm}, we can write the two point correlations of Eq. \ref{eq:twopointcorr}, in terms of the two point correlation function of the primordial density perturbations $\langle\delta(\vec{k})\delta^{*}(\vec{k'})\rangle$, i.e. the primordial power spectrum $P(\vec{k})$. In this paper, our primary objective is to consider an inhomogeneous primordial power
spectrum model and an anisotropic one which would give $l$ and $l\pm1$ correlation explaining the hemispherical anisotropy. In real space, the two point
correlation function of the density fluctuations $\tilde{\delta}(\vec{x})$ can be expressed as,
\begin{equation}
F(\vec{R},\vec{X})=\langle\tilde{\delta}\left(\vec{x}\right)\tilde{\delta}\left(\vec{x}^{\,\prime}\right)\rangle,
\end{equation}
where $\vec{R}=\vec{x}-\vec{x}^{\,\prime}$ and $\vec{X}=\left(\vec{x}+\vec{x}^{\,\prime}\right)/2$. If $F(\vec{R},\vec{X})$ depends only on the magnitude $R\equiv|\vec{R}|$, then in the Fourier space, the two point correlation of $\delta(\vec{k})$ lead to the standard isotropic power spectrum
\[
P_{iso}(k)=\frac{1}{4\pi k^{3}}A_{s}\left(\frac{k}{k_{*}}\right)^{n_{s}-1} \, .
\]
For an anisotropic model $F(\vec{R},\vec{X})$ depends on both the magnitude and 
direction of $\vec{R}$ but is independent of $\vec{X}$ and for an inhomogeneous 
model $F(\vec{R},\vec{X})$ can't be independent of $\vec{X}$. Next we give details about the two models.

\subsection{Inhomogeneous Model}
Following \citet{Carroll2010} and \citet{Pranati2014}, we consider the real space two point correlation function to 
have a mild dependence on $\vec{X}$ and propose the following 
inhomogeneous model of $F\left(\vec{R},\vec{X}\right)$: 
\begin{equation}
F\left(\vec{R},\vec{X}\right)=f_{1}(R)+\sin\left(\vec{\lambda}\cdot\frac{\vec{X}}{\tau_{0}}+\delta\right)f_{2}(R),
\label{eq:Inhomo_real_space}
\end{equation}
where $\vec\lambda$ and $\delta$ are the inhomogeneous parameter and phase
respectively. 
The first term on the right is the standard isotropic and homogeneous two point correlator.
Since $\vec{X}$ has the dimensions of length, we divide it by the present conformal time\footnote{One could use some other scale but since the conformal
time is related to the size of the universe, it is reasonable scale to use.} $\tau_{0}$ in order to make it dimensionless. This model assumes an inhomogeneity in the direction $\vec\lambda$, centered around some point whose position is fixed 
by the phase $\delta$. 
The amplitude of inhomogeneity is parametrized by $f_2(R)$. We expect the
absolute magnitude of $f_2(R)$ to be small. Furthermore we expect
this term to only show mild oscillations and, hence, 
the absolute magnitude of $\vec \lambda$
should not be large compared to unity. 
This model differs somewhat from that proposed in \citet{Pranati2014} where the authors 
expanded the inhomogeneous term in powers of $\vec \lambda\cdot\vec X$ and kept 
only the leading order term. Both the models lead to correlations between 
multipoles $l$ and $l+1$. However they differ in the nature of their higher order
correlations. The sinusoidal parametrization is useful since it damps the 
inhomogeneous term for large values of $\vec X$. This isn't the case for the model used in \citet{Pranati2014}. For non-zero values of the parameter $\delta$
in Eq. \ref{eq:Inhomo_real_space}
 we obtain correlations between multipoles $l$ and $l+2$ also. 
There is currently no evidence for such correlations in data.  
Hence, for simplicity, we set $\delta=0$ in our analysis.

The Fourier transform of the real space two point correlator gives
\begin{eqnarray}
\left\langle \delta(\vec{k})\delta^{*}(\vec{k}^{\prime})\right\rangle  & = & P_{iso}(k)\delta^{3}\left(\vec{k}-\vec{k}^{\prime}\right) -\frac{i}{2}g(k_+) \nonumber \\
 & \times &    \left[\delta\left(\vec{k}-\vec{k}^{\prime}+\frac{\vec{\lambda}}{\tau_{0}}\right)-\delta\left(\vec{k}-\vec{k}^{\prime}-\frac{\vec{\lambda}}{\tau_{0}}\right)\right]
\label{eq:Inhomo_fourier_space}
\end{eqnarray}
where 
\[
g\left(k_{+}\right)=\int\frac{d^{3}{R}}{(2\pi)^{3}}\exp\left(i\left(\vec{k}+\vec{k}^{\prime}\right)\cdot\frac{\vec{R}}{2}\right)f_{2}(R).
\]
and $\vec k_+=(\vec k+\vec k^\prime)/2$.
 In this model, the second term of
Eq. \ref{eq:Inhomo_fourier_space} produces power asymmetry. We parametrize this function as follows:
\begin{equation}
g(k)=g_{0}P_{iso}(k)(k\tau_{0})^{-\alpha},\label{eq:g_parametrization}
\end{equation}
where $g_0$ is the amplitude of the inhomogeneous part, $\alpha$ the spectral
index and, as defined earlier, $\tau_{0}$ is the present conformal time.

\subsection{Anisotropic Model}

Following \citet{Pranati2014b} and \citet{Kothari2015}, 
we assume that the correlation function depends on the direction
of $\vec{R}$ through a dot product with the preferred direction unit vector $\hat{\lambda}$. The form of the power spectrum in real space is
\begin{equation}
F\left(\vec{R},\vec{X}\right)=f_{1}\left(R\right)+\hat{\lambda}\cdot\vec{R}f_{2}\left(R\right)\label{eq:Aniso_real_space}
\end{equation}
where $f_{1}\left(R\right)$ is the real isotropic and homogeneous two point 
correlation function and the second term introduces the anisotropy. Taking a
Fourier transform of the anisotropic term we get
\begin{equation}
\langle\delta(\vec{k})\delta^{*}(\vec{k'})\rangle
=\left[P_{iso}(k)+i\hat{k}\cdot\hat{\lambda}g\left(k\right)\right]\delta^{3}\left(\vec{k}-\vec{k}^{\prime}\right),\label{eq:Aniso_fourier_space}
\end{equation}
In this case we use the same parametrization of $g\left(k\right)$ given by Eq. \ref{eq:g_parametrization} which was used for the 
inhomogeneous model.

\subsection{The Correlation Integrals}
We can calculate the correlation function in Eq. \ref{eq:twopointcorr} in terms of transfer function and the power spectrum by using Eq. \ref{eq:alm} for $a_{X,lm}$.
We obtain,
\begin{eqnarray}
\left\langle a_{X,lm}a_{X,l^{\prime}m^{\prime}}^{*}\right\rangle  & = & \left(4\pi T_{0}\right)^{2}(-i)^{l-l^{\prime}}\int d^{3}kd^{3}k^{\prime}\Delta_{X}\left(l,k\right)\Delta_{X}\left(l^{\prime},k^{\prime}\right)\nonumber \\
 &  & \times Y_{lm}^{*}\left(\hat{k}\right)Y_{l^{\prime}m^{\prime}}\left(\hat{k}^{\prime}\right)\left\langle \delta(\vec{k})\delta(\vec{k}^{\prime})\right\rangle \label{eq:diff_TE_ET}
\end{eqnarray}
This correlation can now be written in the following form,
\begin{equation}
\left\langle a_{X,lm}a_{X,l^{\prime}m^{\prime}}^{*}\right\rangle =C_{X,l}\delta_{ll^{\prime}}\delta_{mm^{\prime}}+\delta_{mm^{\prime}}A_{X,model}\left(l,l^{\prime}\right)
\label{eq:AX}
\end{equation}
where the subscript (\textit {model}) refers to the model, i.e. inhomogeneous or 
anisotropic, being used. As explained earlier, 
the $X$ subscript represents the $T$ and $E$ modes. As has been written above, $C_{X,l}$ is just the isotropic correlation. 
Here we are primarily concerned with the quantity $A_{X,model}\left(l,l^{\prime}\right)$ which relates $l$ and $l^{\prime}$, $l\ne l^{\prime}$. 
We next compute this correlation for the two models under consideration.

\subsubsection{Inhomogeneous Model}

We next compute the contribution to the multipole correlations due to the 
inhomogeneous term. The calculation proceeds along the lines described in \citet{Pranati2014}. The relevant quantity, $A_{X,inhomo}$, is defined in Eq. \ref{eq:AX}. It 
can be expressed as
\begin{eqnarray*}
A_{X,inhomo}\left(l,l^{\prime}\right) & = & 8\pi^{2}T_{0}^{2}(-i)^{l-l^{\prime}+1}\int d^{3}kd^{3}k^{\prime}\Delta_{X}\left(l,k\right)\Delta_{X}\left(l^{\prime},k^{\prime}\right)\\
 &  & \times Y_{lm}^{*}\left(\hat{k}\right)Y_{l^{\prime}m{}^{\prime}}\left(\hat{k}^{\prime}\right)g\left(k_{+}\right)\\
 &  & \times\left(\delta^{3}\left(\vec{k}_{-}+\frac{\vec{\lambda}}{\tau_{0}}\right)-\delta^{3}\left(\vec{k}_{-}-\frac{\vec{\lambda}}{\tau_{0}}\right)\right),
\end{eqnarray*}
where $\vec k_-=\vec k-\vec k'$. We express the spherical harmonics $Y_{lm}(\theta,\phi)$ as, 
\begin{equation}
Y_{lm}(\theta,\phi)=\sqrt{\frac{(2l+1)(l-m)!}{4\pi(l+m)!}}e^{im\phi}P_{l}^{m}(\cos\theta).
\end{equation}
where $P_l^m(\cos\theta)$ are the Legendre polynomials. After $\vec{k}^{\prime}$ integration, we can substitute $\vec{k}^{\prime}=\vec{k}\pm\frac{\vec{\lambda}}{\tau_{0}}$. So up to the first order in the inhomogeneous parameter $\vec{\lambda}$, we can expand, $\Delta_{X}\left(l^\prime,k^{\prime}\right)$ and the Legendre polynomials, $P_{l}^{m}\left(\cos\theta_{k^{\prime}}\right)$ as,

\begin{eqnarray}
\Delta_{X}\left(l^\prime,k^{\prime}\right) & \approx &
\Delta_{X}(l^\prime,k)\pm\frac{\lambda}{\tau_{0}}\cos\theta_{k}\frac{d\Delta_{X}(l^\prime,k)}{dk}\\
P_{l}^{m}\left(\cos\theta_{k'}\right) & \approx & P_{l}^{m}(\cos\theta_{k})\pm\frac{\lambda}{k\tau_{0}}\sin^{2}\theta_{k}\frac{dP_{l}^{m}\left(\cos\theta_{k}\right)}{d\cos\theta_{k}}.\\
g\left(k_{+}\right) & \approx & g(k)\pm\frac{\lambda}{2\tau_{0}}\cos\theta_{k}\frac{dg(k)}{dk}.
\end{eqnarray}
Here $\theta_{k}$ is the polar angle of the vector $\vec{k}$. The angular
integration of $A_{inhomo}\left(l,l^{\prime}\right)$
implies correlations between $l$ and $l\pm1$ with $m^{\prime}=m$. Hence for
$l^{\prime}=l+1$ and $m^{\prime}=m$,
we obtain,

\begin{equation}
 A_{X,inhomo}\left(l,l^{\prime}\right)=T_{0}^{2}\left(4\pi\right)^{2}\frac{1}{\tau_{0}}\sqrt{\frac{(l+1)^{2}-m^{2}}{(2l+1)(2l+3)}}\sum_{i}^{3}I_{i}\left(l,l^{\prime}\right)\label{eq:inhomo_integral}
\end{equation}
where the integrals are defined as follows:
\begin{eqnarray}
I_{1}\left(l,l^{\prime}\right) & = & (l+2)\lambda\int_{0}^{\infty}kdk\Delta_{X}\left(l,k\right)\Delta_{X}\left(l^{\prime},k\right)g\left(k\right)\nonumber \\
I_{2}\left(l,l^{\prime}\right) & = & \lambda\int_{0}^{\infty}k^{2}dk\Delta_{X}\left(l,k\right)\frac{d}{dk}\left(\Delta_{X}\left(l^{\prime},k\right)\right)g(k)\label{eq:deri_samasya}\\
I_{3}\left(l,l^{\prime}\right) & = & \frac{\lambda}{2}\int_{0}^{\infty}k^{2}dk\Delta_{X}\left(l,k\right)\Delta_{X}\left(l^{\prime},k\right)\frac{dg(k)}{dk}\nonumber 
\end{eqnarray}
where we have used the transfer function given in Eq. \ref{eq:gen_def_trans_func}. For the inhomogeneous term, we assume the form of $g(k)$ given in Eq. \ref{eq:g_parametrization}. Since the parameter $\lambda$ arises as a multiplicative factor in all integrals, it is convenient to absorb it in a redefinition of $g_0$. Hence we set $\lambda g_0\rightarrow g_0$. The integrals in 
Eq. \ref{eq:deri_samasya} are evaluated using a suitably modified version of CAMB.

\subsubsection{Anisotropic}

We next compute the contribution to the multipole correlations due to the 
anisotropic term. We follow the procedure used in \citet{Pranati2014b} for our analysis. Here the relevant quantity is $A_{X,aniso}$ which is given by:
\begin{eqnarray}
A_{X,aniso}\left(l,l^{\prime}\right) & = & (-i)^{l-l^{\prime}+1}(4\pi)^{2}T_{0}^{2}\xi_{lm;l^{\prime}m^{\prime}}^{0}\nonumber \\
 &  & \times\int_{0}^{\infty}k^{2}dk\left(k\right)\Delta_{X}\left(l,k\right)\Delta_{X}\left(l^{\prime},k\right)g\left(k\right)
\end{eqnarray}
where 
\[
\xi_{lm;l^{\prime}m^{\prime}}^{0}=\delta_{mm^{\prime}}\left[\delta_{l^{\prime},l+1}\sqrt{\frac{\left(l+1\right)^{2}-m^{2}}{\left(2l+1\right)\left(2l+3\right)}}+\delta_{l^{\prime},l-1}\sqrt{\frac{l^{2}-m^{2}}{\left(2l+1\right)\left(2l-1\right)}}\right].
\]
Here we've used
\begin{eqnarray*}
\int Y_{lm}^{*}\left(\theta,\phi\right)Y_{l^{\prime}m^{\prime}}\left(\theta,\phi\right)\cos\theta d\Omega & =\\
\delta_{mm^{\prime}}\left[\sqrt{\frac{\left(l+1\right)^{2}-m^{2}}{\left(2l+1\right)\left(2l+3\right)}}\delta_{l^{\prime},l+1}+\sqrt{\frac{l^{2}-m^{2}}{\left(2l+1\right)\left(2l+3\right)}}\delta_{l^{\prime},l-1}\right].
\end{eqnarray*}
We are interested in only $l,l+1$ correlations and by using Eq. \ref{eq:gen_def_trans_func} we obtain,

\begin{eqnarray}
A_{X,aniso}\left(l,l^{\prime}\right) & = & \delta_{mm^{\prime}}\delta_{l^{\prime},l+1}(4\pi)^{2}T_{0}^{2}\sqrt{\frac{\left(l+1\right)^{2}-m^{2}}{\left(2l+1\right)\left(2l+3\right)}}\nonumber \\
 &  & \times\int_{0}^{\infty}k^{2}dk\Delta_{X}\left(l,k\right)\Delta_{X}\left(l^{\prime},k\right)g\left(k\right)\label{eq:aniso_integral}
\end{eqnarray}
This integral is evaluated by making suitable changes in the standard Boltzmann
code, CAMB \citep{Lewis2000}.

\section{Data Analysis}
Our data analysis procedure closely follows the approach taken in \citet{Pranati2014}, where the authors 
have done a detailed analysis of the dipole modulation term. Here we 
use the high resolution WAMP-ILC 9 year \citep{Bennett2013} and PLANCK's SMICA 
\citep{Planck14} data. 
For masking the galactic portion we use KQ85 for ILC and CMB-union mask for SMICA.
We use full sky maps for computing correlations between different multipoles.
The full sky maps are generated by filling the masked portions of real maps by 
simulated random isotropic CMBR data. The mask boundaries are suitably
smoothed by downgrading the high resolution maps to a lower resolution after
applying suitable Gaussian beam, as explained in \citet{Pranati2014}. For the case of the 
WMAP data the contribution due to detector noise is also included while generating the random map. 
However contribution due to noise isn't included for the case of
SMICA, since the corresponding noise files are too bulky. For multipole range
$2\leq l \leq 64$ the detector noise gives a relatively small contribution 
\citep{Pranati2014}.
Hence the error induced by neglecting the noise contribution in the case of 
PLANCK data in the above range is expected to be small.

The dipole modulation amplitude parameter is estimated from the downgraded low
resolution maps by computing the statistic $S_H(L)$, as defined below. We had seen in section \ref{Sec:Theory} that the dipole modulation term in the temperature
field, Eq. \ref{eq:dipole_mod}, leads to a non-zero correlation between $l$ and
$l+1$ multipoles. Hence we define a correlation function, $C_{l,l+1}$ as, 
\begin{equation}
C_{l,l+1}=\frac{l(l+1)}{2l+1}\sum_{m=-l}^{l}\left\langle a_{lm}a_{l+1,m}^{*}\right\rangle ,\label{eq:corrl_l+1}
\end{equation}
where $\left\langle a_{lm}a_{l+1,m}^{*}\right\rangle $ varies with $m$ and 
is averaged over the range of $m$ for every $l$. The correlation is being 
normalized by the standard $l(l+1)$ factor. Since the above correlator shows large
fluctuations for individual multipoles we define a statistic $S_{H}(L)$ by summing
over a range of 21 multipoles in three or six bins,
\begin{equation}
S_{H}(L)=\sum_{l=l_{min}}^{l_{max}}C_{l,l+1}.\label{eq:statistic}
\end{equation}
where $l_{min}$ is the starting value for a bin and $l_{max}=l_{min}+20$ starting from $l=2$. In \citet{Pranati2014}, 
the authors confined their analysis to three bins,
$2\le l\le  22$,  $23\le l\le  43$ and $44\le l\le  64$. The downgraded low
resolution maps were chosen to have $N_{side}=32$ for that analysis. Here
we extend this analysis to a large range of multipoles with $l\le 127$ with the 
corresponding maps of $N_{side}=64$. For the three bins with $65\le l\le  85$,  $86\le l\le  106$ and
$107\le l\le  127$ we use only the WMAP data since
contribution due to noise is expected to be significant in this range of multipoles.

We next describe our procedure for estimating the parameter $A$ in each of the bins. We first compute the statistic 
$S_{H}(L)$ in the selected bin from the full sky
CMBR map for which the masked portions have been filled by randomly generated
isotropic data, as described above. This estimate of $S_{H}(L)$ is expected to be
biased due to the filling of masked portions by random isotropic data. We correct this bias by simulations.
We generate 1000 full sky simulated maps which have same characteristics as the real map, including the dipole modulation. 
The details are given in \citet{Pranati2014}. Here 
we briefly review the main steps in generating simulated data. We first generate
a full sky realization of a random isotropic CMBR map. We multiply this map with
the dipole modulation factor, $(1+A\hat\lambda\cdot\hat n)$, where the parameters,
$A$ and $\hat\lambda$, are set equal to the best fit values obtained by the 
hemispherical analysis of the WMAP five year map \citep{Hoftuft2009}. 
We next apply the same mask on the simulated map as was used for the case of real
data. The resulting masked regions are filled with randomly generated isotropic 
CMBR data.
This process generates a full sky realization of a map which has characteristics
similar to that of a real map including the dipole modulation. For each value
of the input parameter $A$ used to generate this map, we generate 1000 realizations and estimate 
the corresponding statistic $S_{H}(L)$ by taking the
average over these maps.
The standard deviation of $S_H(L)$ over these 1000 maps gives an estimate of the
error in the statistic. The best fit value of $A$ is obtained by matching the 
statistic for simulated data with that obtained by real data. The procedure
also gives an estimate of the error in $A$. This value of $A$ can be used in order
to determine the bias corrected estimate of $S_H(L)$ for data. We do this by
generating a full sky realization of the CMBR map, along with the dipole 
modulation corresponding to the best fit value of $A$. The statistic computed
from the resulting map is termed $S_H^{data}(L)$.

The parameters of the theoretical model are computed by fitting the data 
statistic, $S_H^{{data}}(L)$. The theoretical estimate of the statistic, $S_{H}^{{theory}}$, 
for a particular set of parameters $(\alpha,g_0)$ is obtained my making suitable modifications to the standard Boltzmann code,
CAMB, using the cosmological parameters determined by \citet{Planck14}.
We use CAMB to compute the quantities, $A_{X,{inhomo}}$ and $A_{X,{aniso}}$, which are used to compute the statistic, $S_{H}^{{theory}}$. 
The parameters of the two models are estimated by performing a $\chi^{2}$ minimization. The statistic $\chi^{2}$ is defined as
\begin{equation}
\chi^{2}=\sum_{{Bin}}\left(\frac{S_{H}^{{data}}-S_{H}^{{theory}}\left(\alpha,g_{0}\right)}{\delta S_{H}^{{data}}}\right)^{2}.
\end{equation}
In determining the model parameters, we first set $\alpha=0$ and determine the 
best fit value of $g_0$. Next we determine the best fit values of both parameters.

\begin{figure}
\includegraphics[scale=0.2]{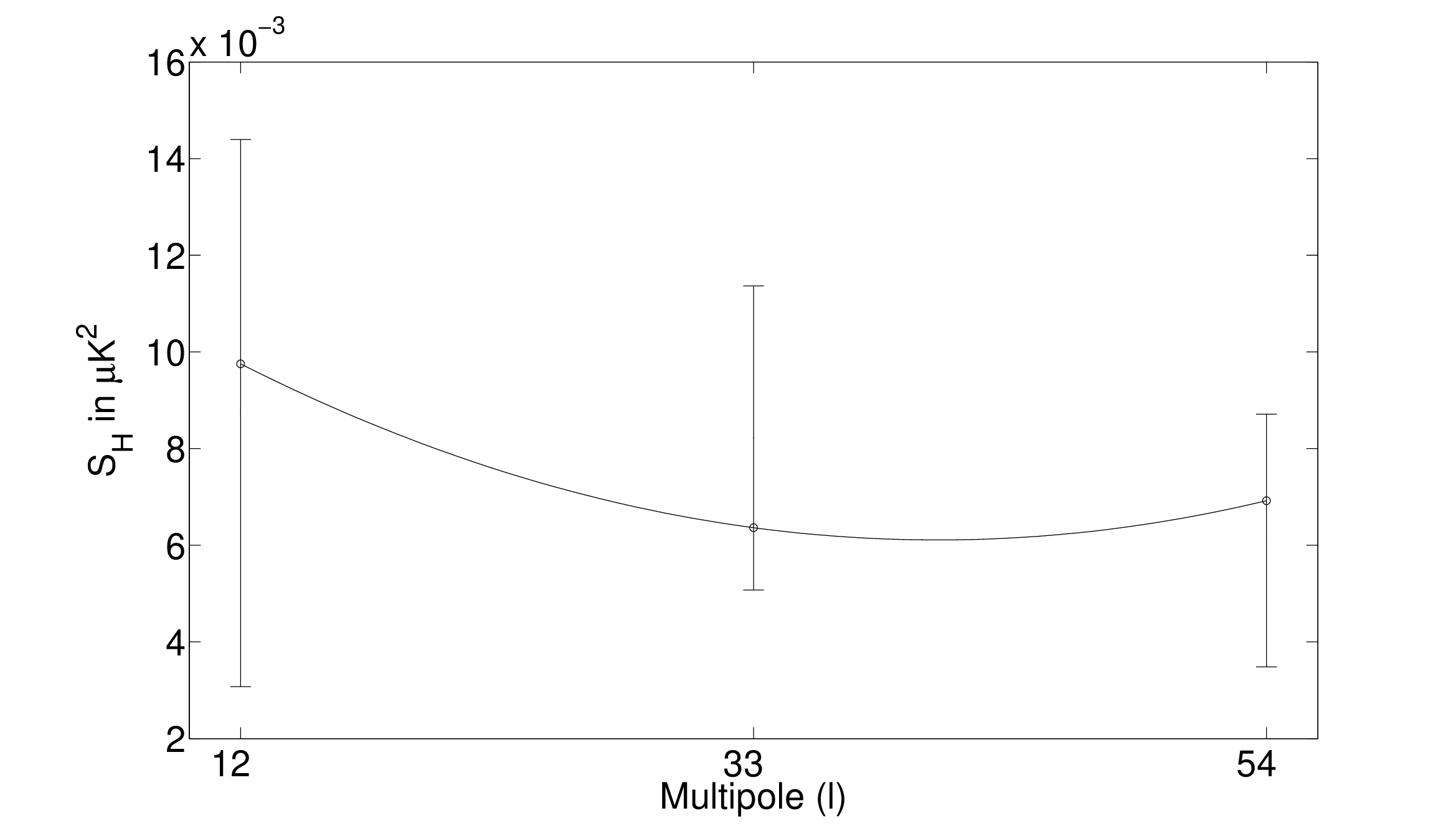} 
\includegraphics[scale=0.2]{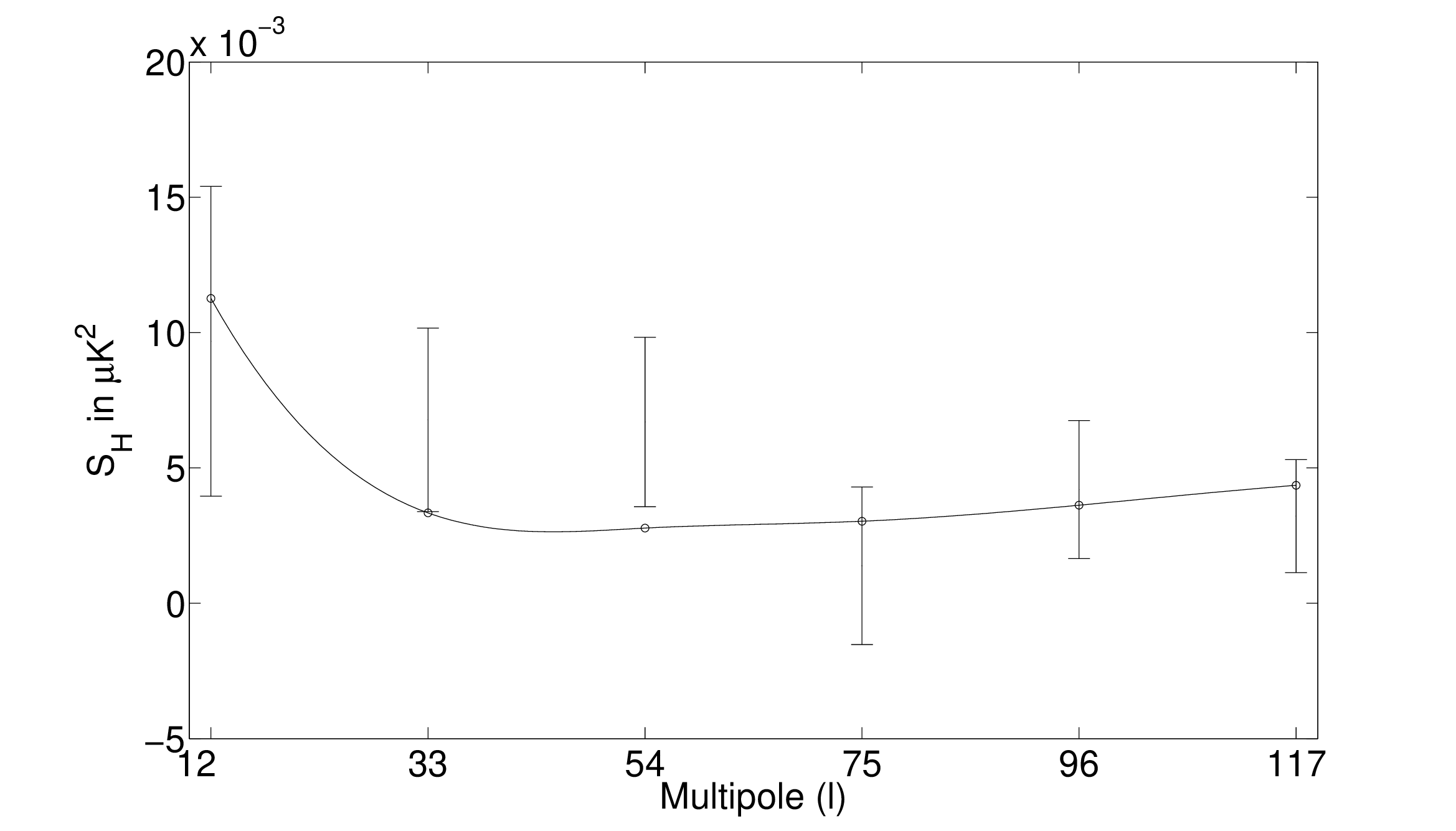} 
  \caption{The fit to observed data for three bins (upper graph) and six bins (lower graph) for WILC9 map.}    
  \label{fig:fit}
\end{figure}

The value for the statistic was calculated independently with WMAP 9 year data
and the PLANCK data. We use the ILC map provided by WMAP, henceforth called 
WILC9, and the SMICA map provided by PLANCK. We use the values of $S_{H}^{{data}}$ 
that were calculated in \citet{Pranati2014} for the case of three bins. Here we also extend this analysis to a larger range of multipoles.
These values are provided in Table \ref{tab:Statistic-from-Data} for 3 bins and
Table \ref{tab:Values-for-statistic} for 6 bins. 
The resulting fits for both the cases are shown in Fig. \ref{fig:fit}. 
 The direction parameters have
been set to be equal to those obtained by hemispherical analysis of the WMAP five
year data in \citet{Hoftuft2009}. These match closely with the parameters extracted in \citet{Pranati2014}
by studying correlations between multipoles $l$ and $l+1$ over the multipole range
 $2\le l\le 64$.
 The difference in the values of the statistic in the first three bins in the 
 Table \ref{tab:Values-for-statistic} in comparison to those in Table
 \ref{tab:Statistic-from-Data} is due to the difference in the resolution
 of the map used. The statistic values in Tables \ref{tab:Statistic-from-Data} and 
 \ref{tab:Values-for-statistic} are extracted by downgrading the high resolution
 map to $N_{{side}}=32$ and 64 respectively. In Table \ref{tab:Values-for-statistic} we find that the value of the statistic is significantly lower for the 
 bins corresponding to $l>64$ in comparison to those of lower $l$ bins. Hence we
 find a clear signal that the dipole modulation effect decays beyond $l=64$.
 
 \begin{table}
 \centering
  
 \begin{tabular}{|c|c|c|c|}
\hline 
Maps & $02\to22$  & $23\to43$  & $44\to64$ \tabularnewline
\hline 
WILC9  & $8.74\pm5.66$ & $8.22\pm3.15$ & $6.10\pm2.62$\tabularnewline
SMICA  & $5.36\pm5.49$ & $6.23\pm3.35$ & $6.60\pm3.16$\tabularnewline
\hline 
\end{tabular}
\caption{\label{tab:Statistic-from-Data} The values for data statistic $S_{H}^{{data}}\times10^{3}$ $\left({m}{K}^{2}\right)$ in three bins using the WMAP nine year (WILC9) and PLANCK (SMICA) data. The direction
  parameters have been set equal to $(l,b)=(224^{\circ},-22^{\circ})\pm24^{\circ}$ \citep{Hoftuft2009}}
\end{table}
 
\begin{table*}
 \centering
 
\begin{tabular}{|c|c|c|c|c|c|}
\hline 
$02\to22$  & $23\to43$  & $44\to64$  & $65\to85$ & $86\to106$ & $107\to127$\tabularnewline
\hline 
$9.68\pm5.72$ & $6.78\pm3.39$ & $6.70\pm3.13$ & $1.38\pm2.91$ & $4.20\pm2.55$ & $3.22\pm2.08$\tabularnewline
\hline
\end{tabular}
\caption{\label{tab:Values-for-statistic} The values for statistic $S_{H}^{{data}}\times10^{3}$
$\left({m}{K}^{2}\right)$ in 6 bins from WMAP (WILC9).}
\end{table*}

\section{Results}

The values of the model parameters obtained by the $\chi^2$ minimization are 
given in Table \ref{tab:Fitting-Parameters}. These deviate from the values 
extracted in \citet{Pranati2014,Pranati2014b} where only the contribution due to the Sachs-Wolfe effect was
included.
We find that for the case of three bins $\chi^2$ is relatively small. Hence the 
model provides a good fit to data. In fact the $\chi^2$ value is too small for the case
of the SMICA map if we allow $\alpha\ne 0$. This is probably due to the fact that
we only have three data points to fit and the error in each is relatively large. The fit
is not found to be as good for the case of six bins. However $\chi^2$ per degree
of freedom is still less than one and hence a pure power law model for either
the inhomogeneous or anisotropic model is acceptable.

\begin{figure*}
\begin{centering}
 \includegraphics[scale=0.43]{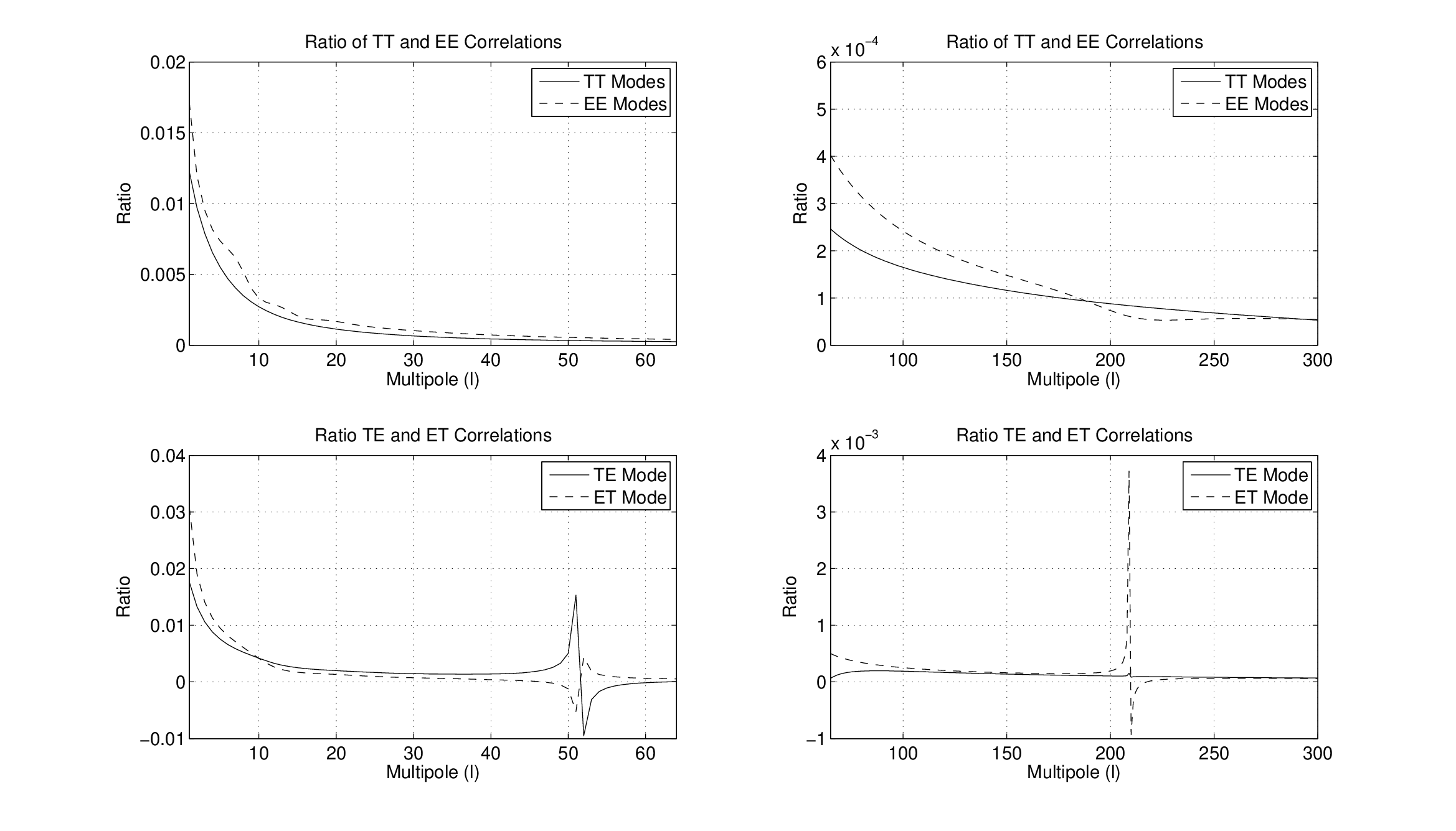}
\end{centering}
\caption{The predicted ratio, Eq. \ref{eq:ratio}, for the case of the 
inhomogeneous model for the EE, TE and ET correlations as a function of the 
multipole $l$. The TT correlations are also shown. The parameter values used
for these plots are obtained by fitting the inhomogeneous model to the SMICA
map for the case of three bins. The left and right plots are for the range
$l\in\left\{ 2,3,\ldots64\right\} $ and $l\in\left\{ 65,66,\ldots300\right\}$ respectively. 
The spikes in ET and TE correlations arise because the power $C_l^{TE}$ crosses zero at the corresponding values.}
\label{fig:Inhomogeneous_part1}
\end{figure*}

\begin{figure*}
\begin{centering}
 \includegraphics[scale=0.43]{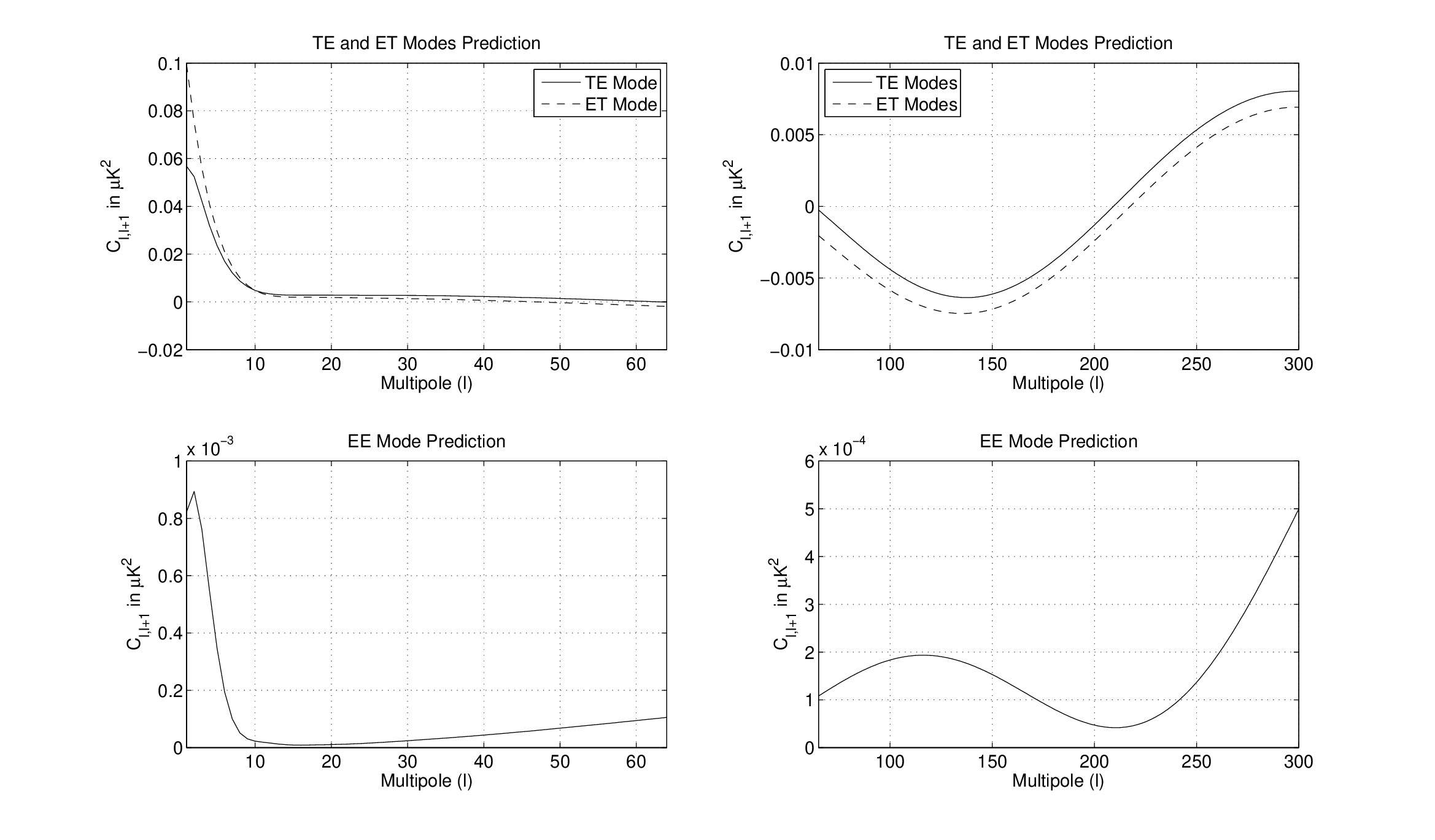}
\end{centering}
\caption{The predicted correlations, $C_{l,l+1}$, as a function of the
multipole $l$ for the TE, ET and EE modes for the inhomogeneous model.}
\label{fig:Inhomogeneous_part2}
\end{figure*}

\begin{figure*}
\begin{centering}
 \includegraphics[scale=0.43]{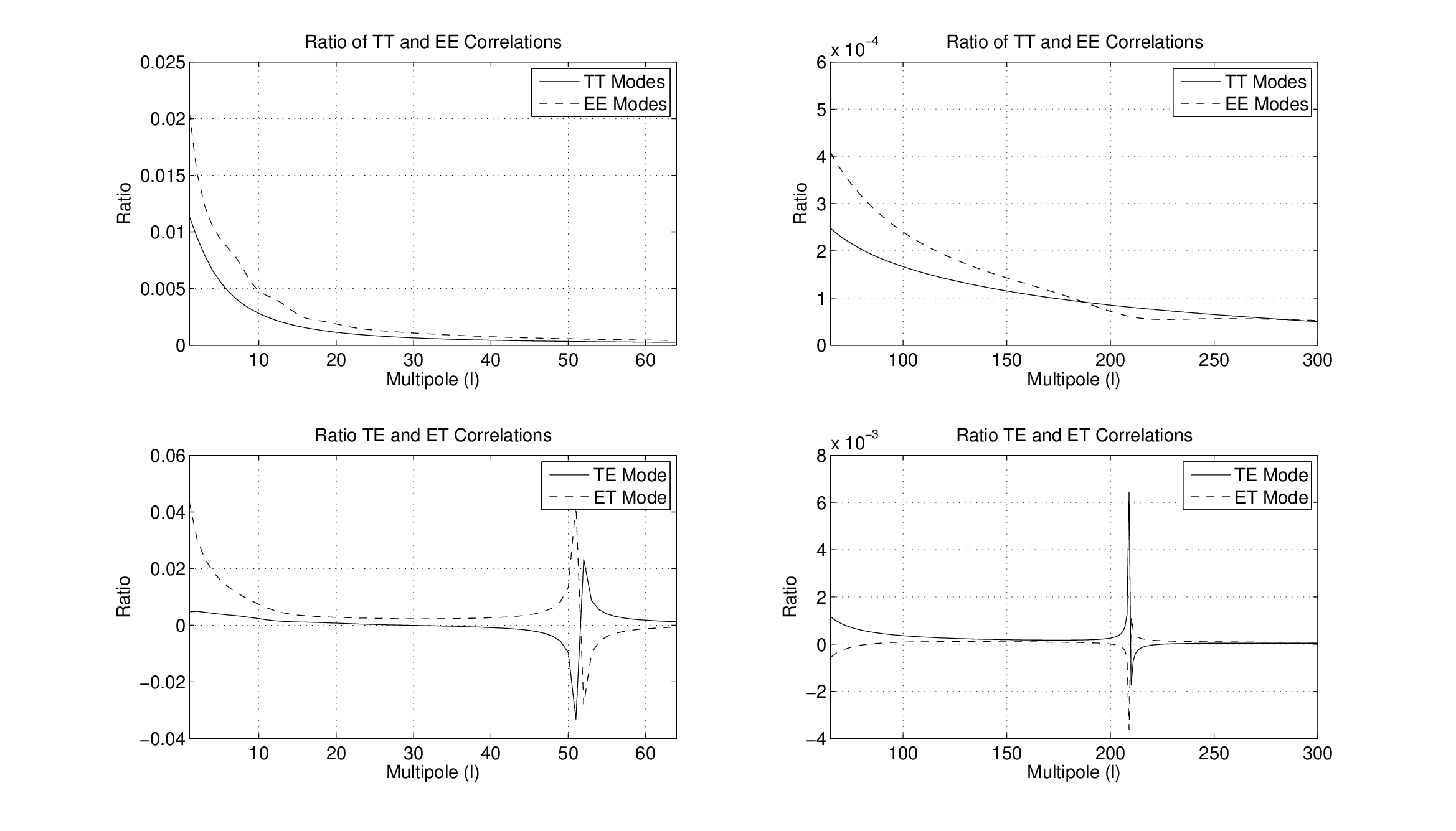}
\end{centering}
\caption{The predicted ratio, Eq. \ref{eq:ratio}, for the case of the 
anisotropic model for the EE, TE and ET correlations as a function of the 
multipole $l$. The TT correlations are also shown. The parameter values used
for these plots are obtained by fitting the anisotropic model to the SMICA
map for the case of three bins.}
\label{fig:Anisotropic_part1}
\end{figure*}

\begin{figure*}
\begin{centering}
 \includegraphics[scale=0.43]{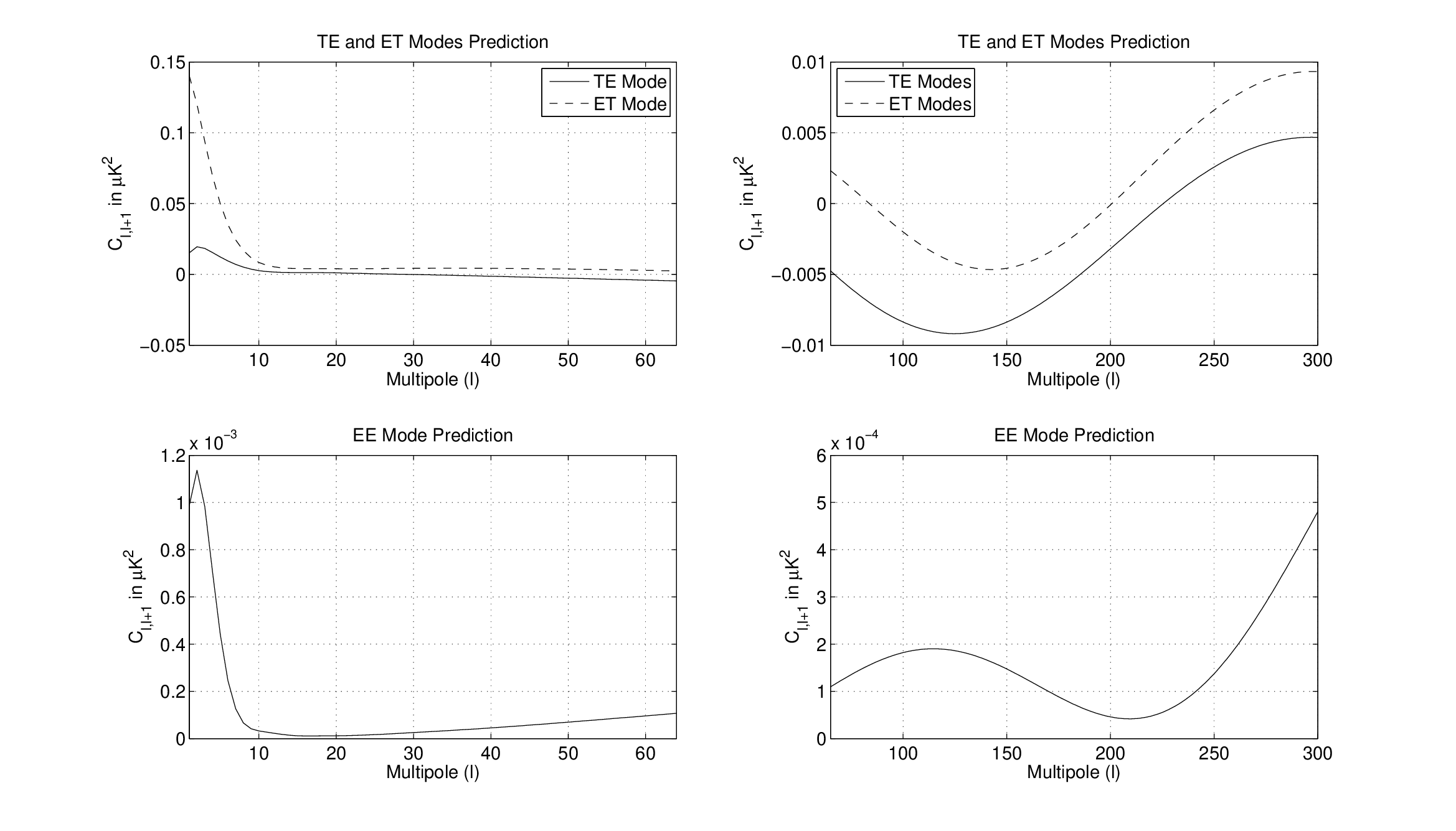}
\end{centering}
\caption{The predicted correlations, $C_{l,l+1}$, as a function of the
multipole $l$ for the TE, ET and EE modes for the anisotropic model.}
\label{fig:Anisotropic_part2}
\end{figure*}

The estimated parameters can be used to predict the signal for the case of 
polarization data. We use the parameters extracted for the case of the SMICA map
in giving our predictions. In this case only data in the three bins, $2\le l\le  22$, $23\le l\le  43$ and $44\le l\le  64$ is used in determining the model
parameters.
These provide a somewhat conservative estimate of the dipole modulation. The 
predictions obtained by using the WILC9 parameters are similar to those obtained 
by using SMICA. However the amplitude in this case is found to be systematically
larger by about 15\%. Our results for TT, EE, TE and ET correlations for the case
of the inhomogeneous model are shown in Figs. \ref{fig:Inhomogeneous_part1} and 
\ref{fig:Inhomogeneous_part2}. In Fig. \ref{fig:Inhomogeneous_part1} we plot the
ratio, defined as,
\begin{equation}
\text{ratio} = {C_{l,l+1}\over C_l}
\label{eq:ratio}
\end{equation}
where $C_l$ is the standard power and $C_{l,l+1}$ is the correlation between $l$
and $l+1$ multipoles, defined in Eq. \ref{eq:corrl_l+1}. In Fig. 
\ref{fig:Inhomogeneous_part2} we show the predicted value of $C_{l,l+1}$ for TE, 
ET and EE modes for the inhomogeneous model. The corresponding values for the
anisotropic model are shown in Figs. \ref{fig:Anisotropic_part1} and \ref{fig:Anisotropic_part2}. We find that the ratio for the TT mode is 
approximately equal to 0.015 at $l=2$ and decays for larger values of $l$. This 
is true for both the models. The ratio for the EE correlations follow the TT
mode closely. The results for the case of $l\le 64$ are shown in the left panels in 
these figures whereas those for larger $l$ values are shown in the right panels.
Since the three bin fit may not be reliable over the larger range of $l$ values
the results for $l>64$ may be treated as qualitative estimates. However the
results for $l\le 64$ are quantitatively reliable and may be compared with data
in order to determine the validity of the inhomogeneous or anisotropic model.

\begin{table*}
\centering
\begin{tabular}{|c|c|c|c|c|c|c|c|}
\hline 
\textbf{No of Bins} & \textbf{Maps} & \textbf{Model} & \multicolumn{3}{c|}{$\alpha\ne0$} & \multicolumn{2}{c|}{$\alpha=0$}\tabularnewline
\hline 
 &  &  & $g_{0}$  & $\alpha$  & $\chi^{2}$  & $g_{0}$  & $\chi^{2}$\tabularnewline
\hline 
3 & WILC9 & Inhomogeneous  & $0.91\pm0.24$  & $0.38\pm0.08$  & $0.48$  & $0.18\pm0.05$  & $1.33$\tabularnewline

3 & WILC9 & Anisotropic  & $1.01\pm0.19$  & $0.42\pm0.11$  & $0.50$  & $0.18\pm0.05$  & $1.44$\tabularnewline
\hline 
3 & SMICA & Inhomogeneous  & $0.67\pm0.21$  & $0.32\pm0.07$  & $6.4\times10^{-3}$  & $0.18\pm0.06$  & $0.46$\tabularnewline

3 & SMICA & Anisotropic  & $0.75\pm0.23$  & $0.36\pm0.20$  & $9.5\times10^{-3}$  & $0.17\pm0.05$  & $0.48$\tabularnewline
\hline 
6 & WILC9 & Inhomogeneous  & $3.25\pm0.89$  & $0.87\pm0.08$  & $3.34$  & $0.04\pm0.02$  & $7.91$\tabularnewline

6 & WILC9 & Anisotropic  & $4.13\pm2.81$  & $0.94\pm0.24$  & $3.23$  & $0.04\pm0.01$  & $8.05$\tabularnewline
\hline 
\end{tabular}
\caption{\label{tab:Fitting-Parameters} The model parameters obtained for the
case of 3 and 6 bins for the inhomogeneous and anisotropic models. 
For the case of 3 bins we give results both for WILC9 and SMICA. 
For the case of 6 bins, we give results only for WILC9 for reasons
explained in text.}
\end{table*}

We give predictions separately for the TE and ET correlations since we find that, 
\begin{equation}
\left\langle a_{T;l,m}a_{E;l+1,m}^{*}\right\rangle \ne\left\langle a_{E;l,m}a_{T;l+1,m}^{*}\right\rangle .
\label{eq:difference}
\end{equation}
The difference between the two arises due to contributions from the inhomogeneous
or anisotropic terms and can be easily extracted from Eq.
\ref{eq:diff_TE_ET}.
This difference is particularly interesting in the neighbourhood of $l\approx 52$ where the power $C_l$ of the $TE$ mode crosses zero. Due to this zero in 
the denominator, the ratio 
tends to be very large in the neighbourhood of such points. We find that for the
inhomogeneous model, the $TE$ mode remains positive whereas the $ET$ mode becomes 
negative for $l\approx40$. Hence the ratio for the $TE$ mode is positive (negative) for $l<52$ ($l>52$). The ET mode shows the opposite trend in the 
neighbourhood of $l\approx 52$. This appears to be a rather interesting
signature of the inhomogeneous model which should be testable with the 
polarization data.
Furthermore the anisotropic model
(Fig \ref{fig:Anisotropic_part1}) shows a more pronounced trend which is 
opposite to what is seen for the case of inhomogeneous model. In this case it is
the ratio of the ET mode which stays positive for $l<52$ and becomes negative
for $l>52$ with the TE mode showing the opposite trend. Hence we find a 
qualitatively different behaviour for these two models which might be useful in
order to distinguish between them.

\section{Conclusion}
We have investigated the implications of the hemispherical power anisotropy
within the framework of an inhomogeneous and independently an anisotropic
Universe model. The Universe is assumed to be inhomogeneous and/or anisotropic
at very early pre-inflationary times. This produces a modification of the 
primordial power spectrum which is parametrized in terms of a simple index
$\alpha$ and amplitude $g_0$ of the inhomogeneous or the anisotropic term
by making a fit to the observed data. The resulting parameters are used to make
predictions for the polarization data, i.e. EE, TE and ET correlations between
multipoles corresponding to $m^{\prime}=m$ and $l^{\prime}=l+1$.

In our analysis we mostly emphasize the results for $l\le 64$ for which the 
contribution due to detector noise is negligible. For the case of WMAP data,
however, we also extend our study to larger $l$ values including the 
contribution due to detector noise. We find that the signal of dipole 
modulation is smaller for $l>64$ in comparison to the lower $l$ values. We use 
the theoretical parameters extracted by fitting data for the three bins
corresponding to $l\le 64$ for the case of SMICA map in showing our predictions.
We find that the results for the case of WILC9 map are very similar.
However the amplitude in this case is found to higher by about 15\% 
in comparison to that found in the case of the SMICA map.
We separately show our predictions for $l\le 64$ and for larger $l$ values. For 
$l\le 64$ our predictions are quantitatively reliable whereas for larger $l$ 
values they should be treated as qualitative estimates.

We show the results for the ratio of $C_{l,l+1}$ (which is the correlation 
between $l$ and $l+1$ multipoles) to the power $C_l$. We find that the results
for EE mode follow the TT mode closely. The ratio is found to be roughly around
0.02 for $l=2$ and decays for larger $l$ values. We find that the results for 
the TE mode differ from those of ET mode. The difference is emphasized in the
inequality shown in Eq. \ref{eq:difference}. For larger $l$ values, in the 
neighbourhood of $l\approx 52$, the two modes show qualitatively different
behaviour which may provide a very clean signature of these primordial models.
These predictions can be tested in the new data available from PLANCK. Alternatively it may provide stringent
constraints on these models.

\section*{Acknowledgments}
Rahul Kothari sincerely acknowledges CSIR, New 
Delhi for the award of fellowship during the work.
Some of the results in this paper have been derived using HealPix package
\citep{Gorski2005}. We used standard Boltzmann solver CAMB (http://camb.info/readme.html) for our
theoretical calculations. Finally, we acknowledge the use of PLANCK data
available from NASA LAMBDA site (http://lambda.gsfc.nasa.gov).

\end{document}